\documentclass[journal = nalefd,manuscript=letter]{achemso}

\usepackage{chemformula} 
\usepackage[T1]{fontenc} 
\usepackage{times}
\usepackage{graphicx}
\usepackage{float}

\author{Elisa Riccardi}

\author{Sylvain Massabeau}

\author{Federico Valmorra}

\author{Simon Messelot}

\author{Michael Rosticher}

\author{J\' er\^ome Tignon}

\affiliation[ENSParis]
{Laboratoire de Physique de l'Ecole Normale Sup\' erieure, ENS, Université PSL, CNRS, Sorbonne Université, Université de Paris, F-75005 Paris, France}

\author{Kenji Watanabe}

\affiliation[NIMS1]
{Research Center for Functional Materials, National Institute for Materials Science, Tsukuba, Ibaraki, 305-0047, Japan}

\author{Takashi Taniguchi}

\affiliation[NIMS2]
{International Center for Materials Nanoarchitectonics, National Institute for Materials Science, Tsukuba, Ibaraki, 305-0047, Japan}

\author{Matthieu Delbecq}

\author{Sukhdeep Dhillon}

\author{Robson Ferreira}

\author{S\' ebastien Balibar}

\author{Takis Kontos}

\author{Juliette Mangeney}
\email{juliette.mangeney@phys.ens.fr}

\affiliation[ENSParis]
{Laboratoire de Physique de l'Ecole Normale Sup\' erieure, ENS, Université PSL, CNRS, Sorbonne Université, Université de Paris, F-75005 Paris, France}

\title{Ultrasensitive Photoresponse of Graphene Quantum Dot in the Coulomb Blockade Regime to THz Radiation}


\begin{document}

\newpage
\begin{abstract}
Graphene quantum dots (GQDs) have recently attracted considerable  attention, with appealing properties for terahertz (THz) technology. This includes the demonstration of large thermal bolometric effects in GQDs when illuminated by THz radiation. However, the interaction of THz photons with GQDs in the Coulomb blockade regime - single electron transport regime - remains unexplored. Here, we demonstrate the ultrasensitive photoresponse to THz radiation (from <0.1 to 10 THz) of a hBN-encapsulated GQD in the Coulomb blockade regime at low temperature (170 mK). We show that THz radiation of  $\sim$10 pW provides a photocurrent response in the nanoampere range, resulting from a renormalization of the chemical potential of the GQD of $\sim$0.15 meV. We attribute this photoresponse to an interfacial photogating effect. Furthermore, our analysis reveals the absence of thermal effects, opening new directions in the study of coherent quantum effects at THz frequencies in GQDs. 
\end{abstract}

\newpage


Graphene quantum dots (GQDs) have attracted considerable attention in recent years due to their unique optical and electrical properties that are directly related to their nanoscale structures. The potential of GQDs for the development of new quantum systems has been widely investigated using optical and transport experiments \cite{Guttinger2012,Trauzettel2007,Ponomarenko2008}. For instance, GQDs of a few nanometers in diameter are very appealing for applications in quantum metrology as GQDs can emit a single optical photon at room temperature with high purity and high brightness\cite{Zhao2018}. Further, GDQs of a few tens of nanometers in diameter have been shown to possess very long relaxation times of electronic excitations (60 ns) \cite{Volk2013}, paving the way for spin qubits with long coherence times. Such large GQDs have also become very attractive for the development of new terahertz (THz) emitters and detectors \cite{Sheng2012} as the quantum electron confinement permits characteristic energy level spacing to reach the few millielectron volt (meV) range (i.e. THz spectral range). For instance, the electronic confinement provides a weak transport gap that can prevent undesirable large dark Zener Klein current that is observed in graphene-based photodetectors \cite{Koppens2014,Vicarelli2012}. Moreover, nonradiative Auger recombination processes, which are detrimental for the development of THz lasers \cite{Mitten2015,Gierz2013,Yadav2018}, can be potentially reduced in GQD as a result of limiting the final states available for electron-electron scattering. Recently, large bolometric effects have been demonstrated in nanostructured quantum dot constrictions in epitaxial graphene grown on SiC when illuminated with THz photons. A huge variation of resistance with temperature was obtained, $>430$ M$\Omega$ K$^{-1}$  below $6$ K, owing to electronic confinement \cite{Elfatimi2016}. However, as no gate electrodes were used to isolate the quantum dots from the leads and to control the chemical potential inside the quantum dots, the bolometric effect was limited to THz heating \cite{Fatimy2018}. Although many interesting phenomena can be revealed in single electron transport regime, the investigation of the THz response of GQDs in the Coulomb blockade regime remains elusive.

Here, we report on the photocurrent response of a hBN-encapsulated GQD to THz radiation in Coulomb blockade regime at low temperature (170 mK). We demonstrate that THz radiation (from <0.1 to 10 THz) of  $\sim$10 pW provides a renormalization of the chemical potential of the GQD of $\sim$0.15 meV, leading to a photocurrent in the nanoampere range. We attribute this ultrasensitive photoresponse of the GQD to interfacial photogating effect and evidence the absence of thermal effects in this Coulomb blockade regime. Our experimental condition uniquely permits us to probe low-energy scale physics phenomena in the quantum dot and the interaction with its environment.

\begin{figure}
 \includegraphics [width=12.5cm] {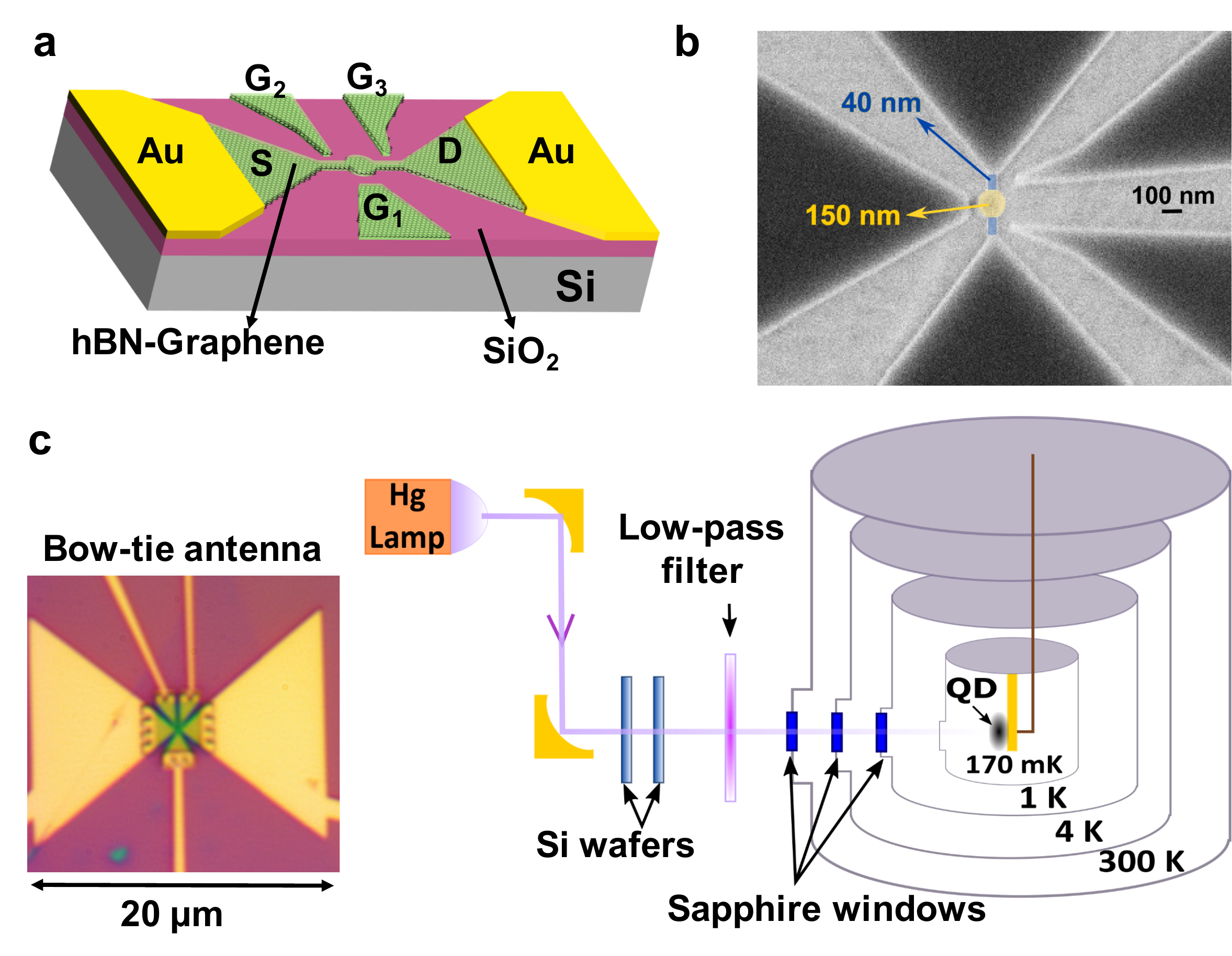}
 \centering
 \caption{\label{Fig:1} (a) Sketch of the devices made of the hBN/graphene/hBN heterostructure deposited on intrinsic silicon substrate with 500 nm thick SiO$_{2}$ on top. We act on the two barriers and on the GQD using three lateral gates $G_1$, $G_2$ and $G_3$ separated from the constrictions and the island by a distance $\sim$50 nm.  We do not apply any backgate voltage using the Si substrate. (b) Scanning electron microscopy (SEM) images of the graphene portion of the device, after the etching of the hBN-graphene-hBN heterostructure. The picture shows the GQD surrounded by the three lateral gates and linked to the source and drain electrodes. (c) Left: Optical image of the connected device with the gold bow-tie antenna. Right: sketch of the experimental setup: the GQD-based sample is placed within a dilution cryostat at 170 mK equipped with three sapphire windows. The incoherent THz radiation emitted by a Hg lamp,  filtered using a low pass filter to frequencies $<10$ THz, is focused on the GQD-based sample. Some high-resistivity silicon wafers are used to attenuate the incident THz radiation.}
 \end{figure}

We investigate a single electron transistor made of a GQD coupled to a bow-tie THz antenna. For achieving quantized energy levels in the range of few meV close to the Dirac point (i.e. in the THz spectral range), we focus on a large GQD with typical diameter of $>$100 nm \cite{Guclu2010}. 
The GQD-based single electron transistor, shown in Fig. \ref{Fig:1}, consists in a large central GQD, with a diameter of $150$ nm, linked to the source and drain electrodes by two narrow constrictions ($40$ nm) and surrounded by three lateral gates ($G_{1}$, $G_{2}$ and $G_{3}$). 
As well as the GQD, all the electrodes and constrictions are made of hBN-encapsulated graphene, deposited on a SiO$_{2}$/Si substrate (see Fig. \ref{Fig:1}(a)). 
Owing to the encapsulation of the graphene layer with two hBN layers, the disorder in the GQD is reduced and the chemical potentials of the graphene leads $\mu_{S,D}$ are expected to be typically within $\pm 25$ meV. 
The graphene device is connected to the gold electrodes and to a THz bow-tie antenna as shown in Fig. \ref{Fig:1}(c).
The whole area of the GQD-based device that includes the whole graphene area (leads and quantum dot) and the metallic bow tie antenna is $20$x$10$ $\mu$m$^{2}$. Further details on the sample fabrication are presented in the Supporting Information (Experimental details).
To perform transport measurements under THz illumination, the GQD-based device is placed within a dilution $^{4}$He-$^{3}$He cryostat with THz optical access (three sapphire windows). The device is cooled to a temperature of $170$ mK. 
The THz light source is a high-pressure mercury plasma arc-discharge lamp. The blackbody radiation emitted by the source is filtered above 10 THz with a low pass filter so that the incident THz photon energies $\hbar\omega $ are  lower than $40$ meV. To focus the THz beam onto the GQD-based device, a pair of parabolic mirrors are placed in front of the fridge window (Fig. \ref{Fig:1}(c)). The incoherent incident radiation is mechanically modulated at a frequency of 330 Hz and we record the DC current with and without illumination (i.e. the mean value of the modulated current) and also the photocurrent, $I_{photo}$, using a lock-in amplifier. 
For optical power-dependence measurements, we insert Si wafers in the THz beam path to attenuate the incident THz power. Without any Si wafer, the estimated THz optical power of the incident THz radiation onto the GQD-based device is $\sim$10 pW.

\begin{figure}[H]
 \includegraphics [width=12cm] {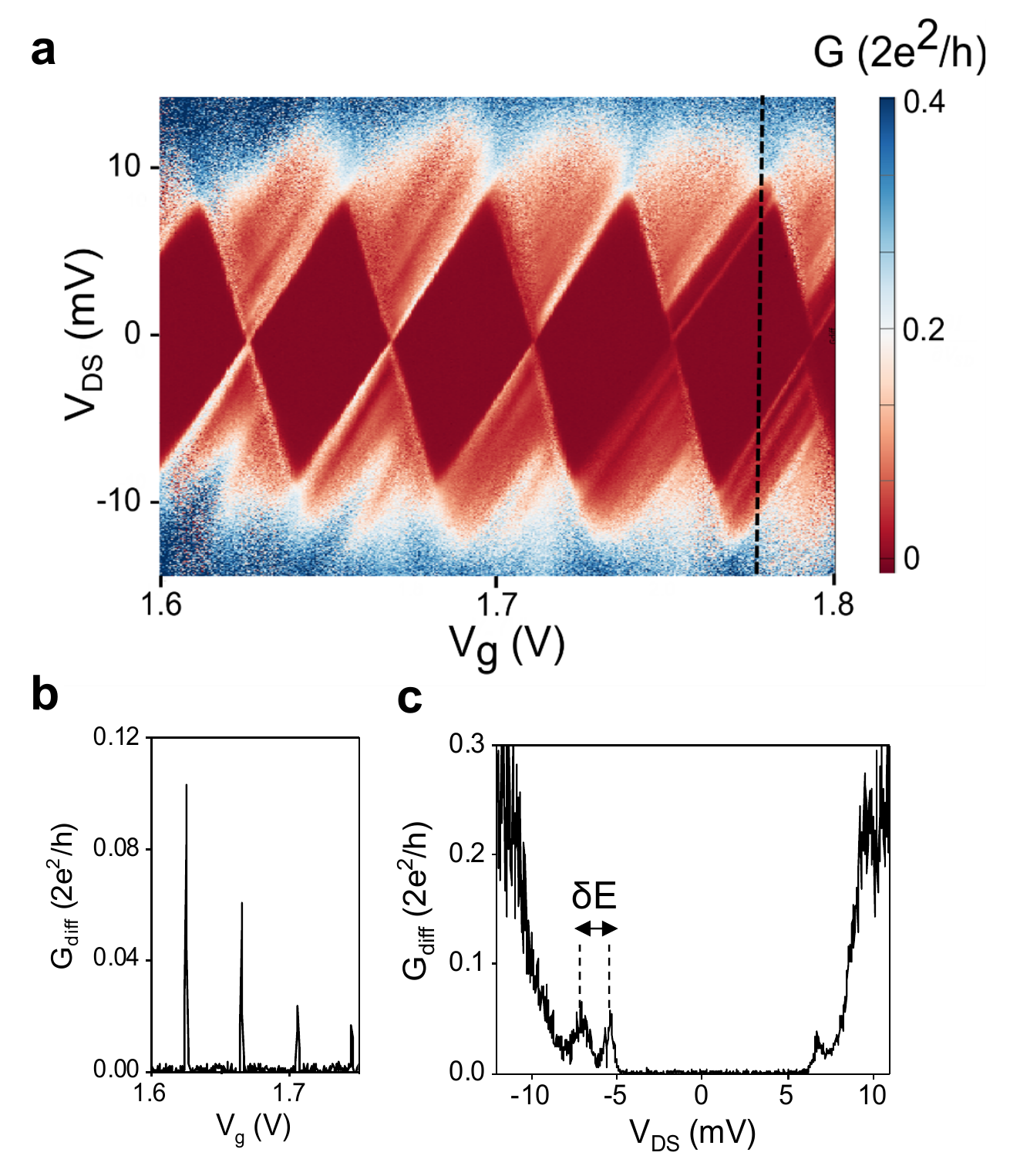}                                           
\centering 
 \caption{\label{Fig:2}(a) Differential conductance G as a function of the source-drain voltage $V_{DS}$ and of the plunger gate voltage $V_{g}$, exhibiting Coulomb diamonds. Multiple resonances parallel to the edges of the diamonds are clearly visible. (b) Differential conductance as a function of $V_{g}$ at $V_{DS}=0 mV$. The two side gates are $V_{G1}=V_{G2}=11.5V$.  (c) Differential conductance as a function of $V_{DS}$ (vertical cut along the dark dashed line in (b)). The width of the diamond allows to estimate the addition energy, while the slope of the diamond edge is proportional to the electronic temperature. }
 \end{figure}

We first perform transport measurements with the fridge's window closed to explore the electron confinement and excitation spectrum of the GQD. The plunger gate $G_{1}$ mainly acts on the chemical potential of the quantum dot, $\mu_{dot}$, and the two side gates $G_{2}$ and $G_{3}$ control the transport through the two constrictions  (see Fig. \ref{Fig:1}(a)). In order for the two constrictions to act as  tunneling barriers, we set them in their respective transport gap at $V_{G2}=V_{G3}=11.5$ V. As observed in Fig. \ref{Fig:2}(b), pronounced Coulomb-blockade peaks are observed in the differential conductance as a function of the plunger gate voltage $V_{g}$,  showing the regime of sequential tunneling where the transport is allowed only if the chemical potential of the source (or drain) crosses one empty level of the quantum dot. 
The Coulomb stability diagram of the GQD-based device, i.e. plots of the differential conductance $G=dI/dV_{DS}$ as a function of  $V_{DS}$ and $V_{g}$, is reported in Fig. \ref{Fig:2}(a). We observe well-defined and stable Coulomb diamonds, which correspond to the ground state of the GQD. The plunger gate voltage $V_{g}$ fills the quantum dot with electrons by moving the energy levels, while the bias voltage $V_{DS}$ shifts the chemical potential of the electron baths of source and drain. The conductance map is not centered at $V_{DS}=0$ V due to the offset voltage of the preamplifier on the drain $V_{0}=-0.35$ mV. The source-drain voltage applied to the GQD is $V_{DS}^{*}=V_{DS}-V_{0}$.  From the half-height of Coulomb diamond, we extract the addition energy needed to add an electron from a lead to the GQD, $E_{add}=9$ meV. This addition energy is expressed as $E_{add}=E_{c}+\delta E$ with $E_{c}=e^{2}/{C_\Sigma}$ as the charging energy, $C_\Sigma$ as the total capacitance of the device, and $\delta E$ as the intrinsic level spacing of the GQD. 
 We also observe quantum confinement effects  as lines parallel to the edges of the diamonds in Fig. \ref{Fig:2}(a). The lines evidence the sequential tunneling of the electrons via the excited states of the GQD. From the vertical cut of the conductance map shown in Fig. \ref{Fig:2}(c), we clearly see  resonances due to the tunneling through the excited states around a region where conductance is strongly suppressed. The energy separation of the excited states gives the single particle level spacing of the GQD, $\delta E=1.7$ meV, corresponding to a frequency spacing of $0.42$ THz. This result confirms that large GQD (of diameter 150 nm) provide discretization of electronic states with energy spacing in the meV range (i.e THz range), as predicted by the literature \cite{Sheng2012,Schnez2009}. The full-width at half-maximum of the excited state resonance at $V_{g}=-5$V is $0.48$ meV corresponding to a thermal broadening of $T=1.8$ K that is roughly consistent with the expected electronic temperature in the GQD under bias. From $E_{add}$ and $\delta E$, we deduce a charging energy $E_{c}=7.3$ meV, which is smaller than the charging energy estimated from a metallic disk model $E_{c} = e^{2}/(4\epsilon_{0}\epsilon_{eff}d)=12$ meV (with $\epsilon_{eff}= (\epsilon_{SiO_{2}} + 1)/2=2.5$) which is in agreement with previous works on GQD-based devices \cite{Guttinger2012}. This deviation is attributed to the capacitive coupling of the GQD to the adjacent gates and leads, which increases the total capacitance $C_\Sigma$. 

\begin{figure}
 \includegraphics [width=12cm] {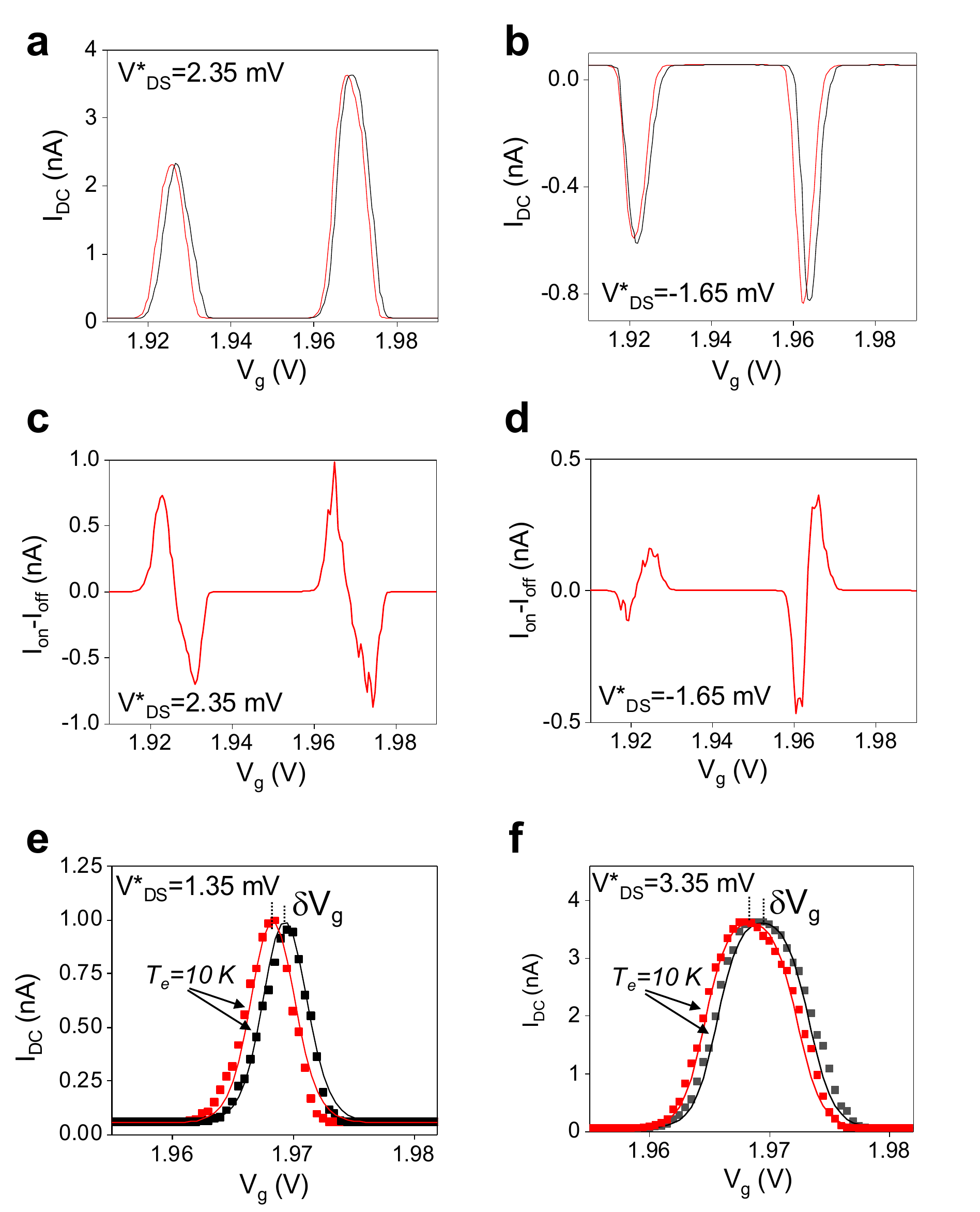}
\centering 
 \caption{\label{Fig:3} (a) DC current Coulomb peaks as a function of $V_{g}$ at $V^*_{DS}= 2.35$ mV (a) and $V^*_{DS}= -1.65$ mV (b) with (red) and without (black) THz illumination. 
We defined $V^*_{DS}$  as the source-drain voltage applied to the GQD after subtracting the preamplifier offset voltage $V_0= -0.35$ mV. 
The difference $I_{on}-I_{off}$ as a function of $V_{g}$ at $V^*_{DS}=2.35$ mV (c) and $V^*_{DS}=-1.65$ mV (d). DC current Coulomb peaks around $V_{g}=1.97$ V at $V^*_{DS}=1.35$ mV (e) and $V^*_{DS}=3.35$ mV (f) with (red) and without (black) THz illumination. Data are reported by diamond symbols, and theoretical fits using equation \eqref{I} are represented by plain lines.}
 \end{figure}

We now turn to the transport investigation of the GQD-based device under THz light illumination. We first focus on the effect of the THz illumination on the current flowing through the GQD at fixed $V_{DS}$. Figures \ref{Fig:3} (a) and (b) show Coulomb blockade peaks in the DC current measured as a function of  $V_{g}$ with (red) and without (black) illumination for $P_{0}=10 pW$ and $V_{DS}=2$ meV and $V_{DS}=-2$ meV, respectively. The magnitude of the Coulomb blockade peaks reach a few nanoamperes. Note that the Coulomb peaks are significantly broadened when opening the fridge's window even without THz illumination. This broadening may be due to the sensitivity of the GQD-based device to the electromagnetic environment. We observe that the magnitude and the shape of the Coulomb current peaks are not affected by the incoming THz photons. However, the current peaks are shifted to lower $V_{g}$ when THz radiation is switched on. The shift, $\delta V_{g}$, is as large as $\sim$1 mV giving rise to a difference $I_{on}-I_{off}$ in the nanoampere range, as shown in \ref{Fig:3}(c) and (d). The $I_{on}-I_{off}$ traces show consistently a bipolar behavior with a positive and a negative peak as a result of the peak current shift to lower $V_{g}$. To get a more quantitative insight in the detection process of the incoming THz photons, we calculate the net current through left and right leads assuming that $\Gamma_{R}, \Gamma_{L}< k_{B}T$ with $\Gamma_{R}, \Gamma_{L}$ representing the tunneling rate between the quantum dot and the left and right leads respectively. Without illumination, the current is given by
\begin{equation}
 I=-eA\lbrack f_{FD}(\alpha_{1}V_{g}-eV_{DS}^{*})-f_{FD}(\alpha_{2}V_{g}-eV_{DS}^{*})\rbrack
\label{I}
\end{equation}
 
with $A=\frac{\Gamma_{L}\Gamma_{R}}{\Gamma_{L}+\Gamma_{R}}$, $\alpha_{1}=\frac{C_{g}}{C_{\sum}-C_{L}}$, $\alpha_{2}=-\frac{C_{g}}{C_{L}}$, $C_{g}$ the capacitance from the GQD to its local gate, $C_{L}$ the capacitance from the GQD to left lead, $C_{\sum}$ the sum of all capacitances on the GQD and $f_{FD}(E)=\frac{1}{e^{E/(k_B T_{e})}+1}$ the Fermi-Dirac distribution. The variables $\alpha_{1}=0.89$ and $\alpha_{2}=0.85$ are extracted from dark current measurements performed with a closed window (see Supplementary Note 1). The dark current peaks around $V_{g}=1.97$ V at $V_{DS}=1$ meV and $V_{DS}=3$ meV without THz illumination (black symbols) are well reproduced using equation \eqref{I} for $T_{e}=10$ K (black lines), as shown in Fig. \ref{Fig:3}(e) and (f) respectively. The parameter $A$ only is adjusted between $V_{DS}=1$ meV and $V_{DS}=3$ meV since the tunneling rates depend on $V_{DS}$. The good agreement between theoretical curves and measured dark DC current validate our description of the current flow through the GQD. By replacing $V_{g}$ with $V_{g}-\delta V_{g}$ in equation \eqref{I}, the DC current peaks under THz light illumination (red symbols) are remarkably well fitted (red lines) without changing any other parameters (see Fig. \ref{Fig:3}(e) and (f)). This indicates that $\delta V_{g}$ is independent of $V_{DS}$. Moreover,  $T_{e}$ is kept constant with and without illumination indicating that the electronic temperature in the GQD is not significantly changed by the incoming THz photons, which is in contrast with previous works on GQD illuminated by THz photons \cite{Elfatimi2016}. This analysis confirms that the THz detection process is fully described by an additional gate voltage $\delta V_{g}$ that leads to a reduction of the GQD chemical potential $\mu_{dot}$ with respect to those of the leads. The responsivity of the GQD-based device reaches $\sim 100$ A/W at $T=170$ mK, demonstrating the high sensitivity for THz photon detection of the GQD-based device at low temperature. 

\begin{figure}[t]
 \includegraphics [width=17cm]  {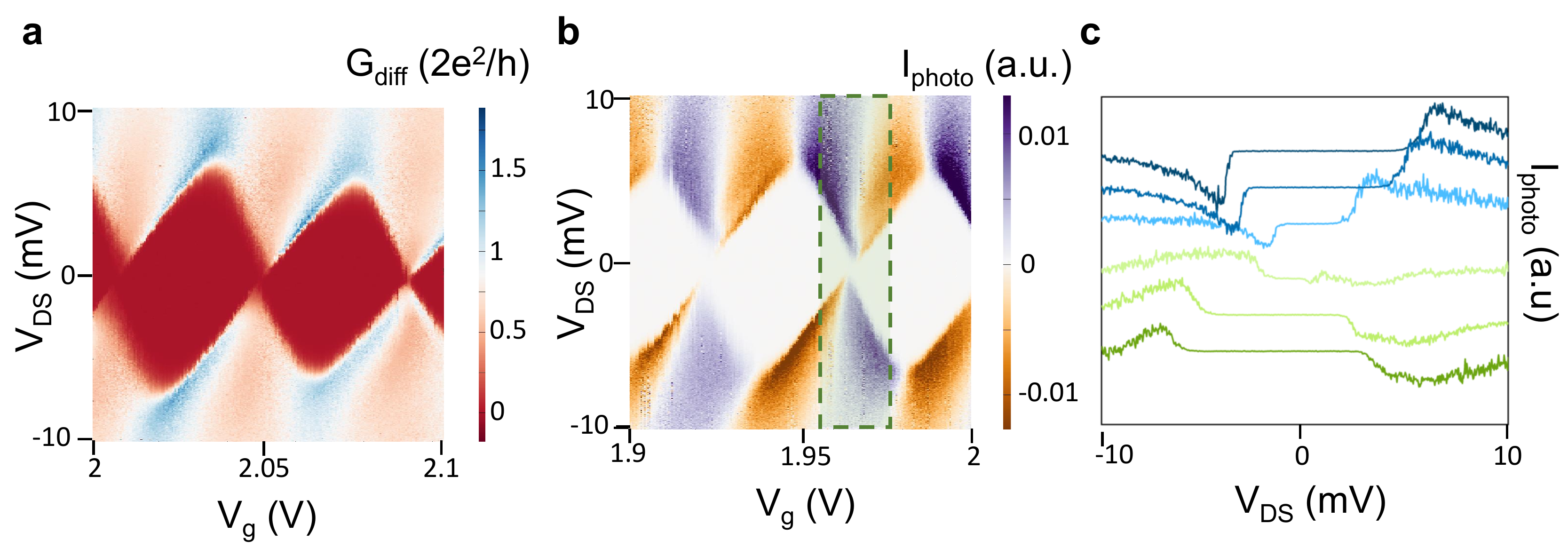}
 \caption{\label{Fig:4} (a) Dark differential conductance G as a function of $V_{DS}$ and $V_{g}$ with fridge's window open. (b) Map of $I_{photo}$ as a function of $V_{DS}$ and $V_{g}$. (c) Vertical cuts of $I_{photo}$ taken in the dashed box of the photocurrent map (b)). The spectra, as a function of $V_{DS}$, exhibit a change in sign when the sign of $V_{DS}$ is changed, and this trend is inverted when the system pass from the left side (green) to the right side (blue) of the Coulomb diamonds. }
 \end{figure}

We further investigate the photoresponse of the GQD-based device as a function of both $V_{DS}$ and $V_{g}$ reported as a $I_{photo}$ map in Fig. \ref{Fig:4}(b) for $P_{0}=10$ pW. For comparison, we also report in Fig. \ref{Fig:4}(a) the differential conductance stability diagram measured without THz illumination but with fridge's window open. The Coulomb diamonds of the dark differential conductance are smaller than those with the window closed (Fig. \ref{Fig:2}(a)), which may be because of the sensitivity of the addition energy and also the capacitance of the GQD-based device to the electromagnetic environment. The dark differential conductance and $I_{photo}$ maps show very similar features such as common diamond-shaped structures. The bipolar features observed only in the $I_{photo}$  map emphasized by the positive and negative slopes of the Coulomb diamonds edges point out significant modification of the resonance conditions induced by THz illumination for charging and discharging the GQD to the respective leads. Thus, the $I_{photo}$ profile in the axis of $V_{DS}$ reflects the tuning, induced by THz light illumination, of the chemical potential in the GQD, $\mu_{dot}$, with respect to those of the leads (Fig. \ref{Fig:4}(c)). 
We attribute the asymmetry of $I_{photo}$ with respect to $V^*_{DS}$  to different capacitances from the GQD to the left and right leads. 
No distortion of the Coulomb diamonds is observed in the $I_{photo}$ map indicating that this chemical potential shift  is independent of $V_{g}$ and $V_{DS}$, i.e., that is constant over the whole charge stability diagram. Thus, illuminating the GQD-based device with incoherent THz photons results in a renormalization of the chemical potential of the GQD, $\mu_{dot}$, relative to those of the source and drain leads. Using the lever arm of $0.15$ extracted from the dark differential conductance map with open fridge window, we deduce that $\delta V_{g}$ of $1 \pm 0.2$ mV corresponds to a $\Delta \mu_{dot}$ of $\sim 0.15 \pm 0.03$ meV. This very low-energy  physical effect is clearly observable here owing to the extremely high-energy resolution provided by the low temperature experiment. Besides, the absence of any additional photocurrent lines induced by THz light into or out of the Coulomb blockaded regions suggests that any intersublevel transition in the GQD or any photon-assisted tunneling effect (electrons exchange photons in the quantum dot) are driven by the incident photons due to the incoherence of the THz radiation.

Let us now discuss the physical mechanism responsible for the renormalization of the chemical potential of the GQD. THz photons, incoming from the incoherent and broadband source, are absorbed by both the whole graphene area (including the quantum dot, the leads and gate electrodes that are made in encapsulated graphene) and by the silicon substrate. Let us first consider if the absorption of THz radiation by the graphene area could be responsible for the $\mu_{dot}$ renormalization. THz photons are preferentially coupled to the leads due to their larger size and their linear dispersion relation (in contrast to the quantum dot) that provides stronger absorption of the broad THz radiation. 
At these frequencies the absorbed energy results in an increase of the electronic temperature and consequently, the chemical potentials of the graphene leads are reduced. \cite{Mics2015} (for details see Supplementary Note 5).
The observed gate renormalization in the stability diagram can then reflect a relative shift between the dot discrete levels and the chemical potential of the graphene leads (see Figure S\ref{Fig:5}(a)). 
 From our analysis described in Supplementary Note 5, we estimate the increase of the electronic temperature for $\Delta \mu_{dot}=0.15 meV$ to be  $\Delta T_{e}>7$ K. 
 However, our  data are qualitatively and quantitatively inconsistent with such temperature increase. Indeed, the current peaks with and without illumination are well reproduced using equation \eqref{I} for a constant temperature $T_{e}=10$ K;  increasing  $T_{e}$ to $>$17 K would significantly broaden the current peak under THz illumination. Moreover, by resolving the heat flow, we estimate $\Delta T_{e}$ in our experimental condition to remain below $0.2$ K. Details of the calculation based on the Wiedemann-Franz law are provided in the Supporting Information (Supplementary Note 4). Consequently, even if a renormalization of the GQD chemical potential is expected from the absorption of the THz photons by the graphene leads, its magnitude would be significantly lower than $0.15$ meV and thus could not account for the large photocurrent response of the GQD-based device. 

Another effect that can be pointed out is the photogating effect. \cite{Shimatani2019,Fang2017} It relies on the absorption of THz photons in the silicon substrate. Absorption of typically $0.2-0.4$ cm$^{-1}$ in the THz spectral range have been recorded in high-resistivity silicon due to shallow impurity absorption (since silicon bandgap energy is high compared to the THz photon energy) \cite{Dai2014,Palik1998,Jiang_2014}. This absorption process creates photoexcited carriers in silicon from the ground states of impurities into the valence and conduction bands \cite{Hubers2001}. This is in contrast with free carrier absorption that does not create free carriers and is extremely low in high-resistivity silicon ($\alpha_{Drude}=0.011cm^{-1}$ due to the low free carrier concentration, $n=10^{11} cm^{-3}$). The photogating effect relies also on the existence of a band bending at the SiO$_{2}$/Si interface due to charges in the oxide (fixed-oxide charge, oxide-trapped charge and mobile ionic charge) and traps at the interface (interface-trapped charge) \cite{Guo2016}. In our case of the silicon substrate with residual N-type doping (see the Supporting Information, Experimental details), the energy bands in silicon bend upward, leading to a triangular potential well for the holes at the interface and a built-in electric field near the interface. The electric field can separate the photoexcited carriers absorbed in silicon, electrons diffuse toward the bulk silicon, and holes accumulate at the interface.\cite{Liu2017,Ogawa2019} 
Holes are moreover trapped in the potential well. This accumulation of photon-generated holes at the SiO$_{2}$/Si  interface provides an additional positive gate voltage by capacitive coupling. In other words, the photogating effect would cause a negative shift in the $I_{DS}$-$V_{G}$ characteristic under laser illumination. This interfacial photogating effect is qualitatively fully consistent with our observations. Indeed, we observe a negative shift of the $I_{DS}$-$V_{G}$ characteristic under THz illumination, independent of $V_{DS}$ and $V_{G}$. In addition, the photocurrent is the difference between the source-drain currents in dark and illumination regime.

\begin{figure} [H]
 \includegraphics [width=10cm] {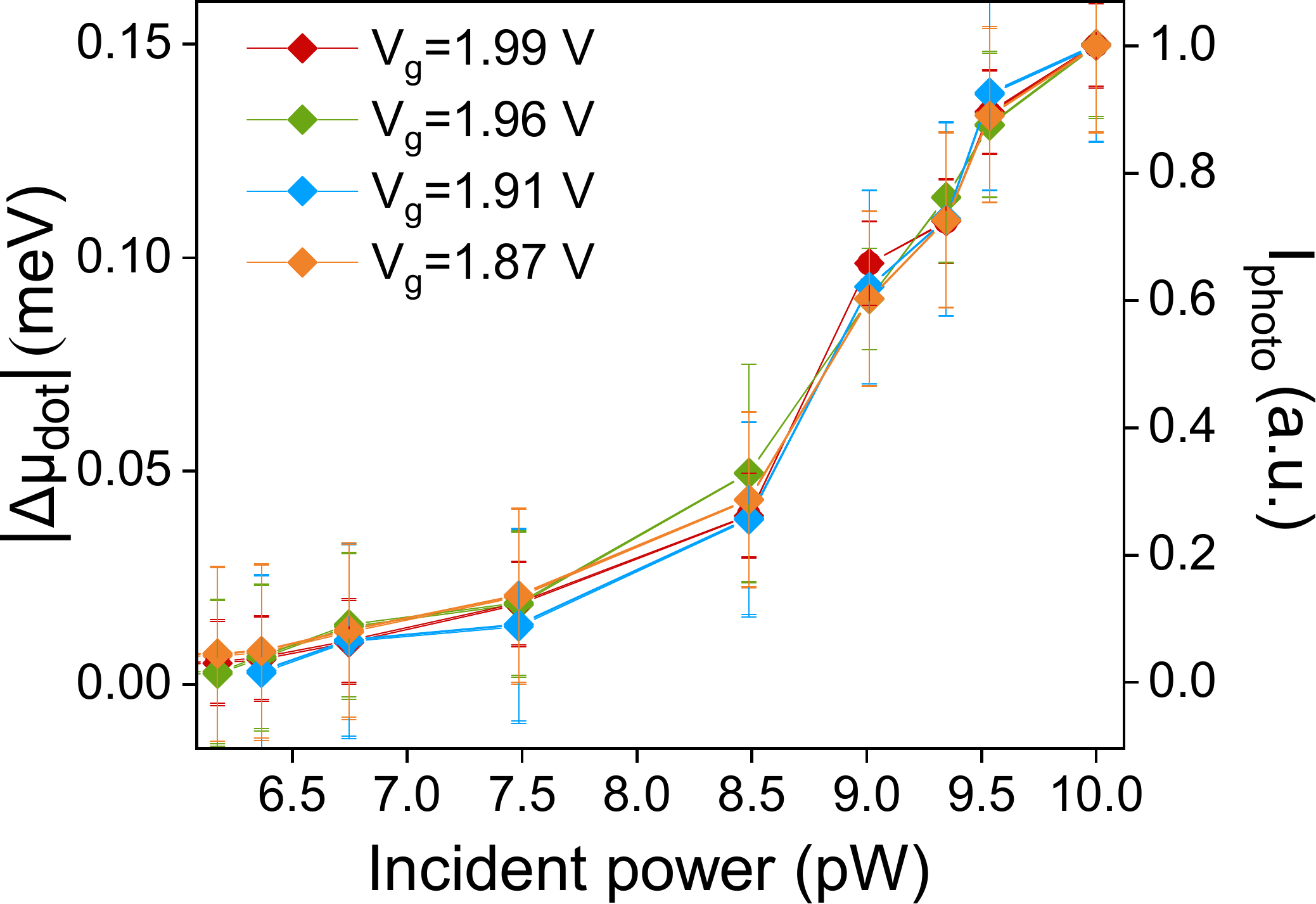}
\centering
 \caption{\label{Fig:5} $\mid$$\Delta \mu_{dot}$$\mid $ and $I_{photo}$ as a function of the incident power $P_{0}$ onto the GQD-based device for $V_{DS}=2$ mV and $V_{g}=1.99$ V; $V_{g}=1.96$ V,$V_{g}=1.91$ V and $V_{g}=1.87$ V. }
 \end{figure}

Several other mechanisms can be responsible for the conversion of absorbed THz photons into a photocurrent in graphene-based devices such as the photovoltaic effect, the photothermoelectric effect and the bolometric effect \cite{Koppens2014}. However, all these mechanisms generate photocurrents or  photovoltages without any shift in gate voltage. Moreover, photothermoelectric and bolometric effects rely on an increase of the electronic temperature that is extremely weak in this work due to the low incident power as discussed previously. Consequently, we attribute the photocurrent originating from the gate voltage shift measured when the GQD-based device is illuminated by THz photons to the interfacial photogating effect. Since the Coulomb blockade current at low temperature is extremely sensitive to the gate voltage, the GQD acts as a very sensitive electrometer operating in the THz spectral range. 
To further investigate the properties of the THz photoresponse, we measure $I_{photo}$ and $\Delta \mu_{dot}$ as a function of $P_{0}$, as displayed in Fig. 5. The same set of measurements is repeated at several positions of the Coulomb diamonds map, i.e., for 4 different values of $V_{g}$. We observe that $I_{photo}$ and $\mid$$\Delta \mu_{dot}$$\mid $ increase monotonously as $P_{0}$ is increasing with a nonlinear dependence. $I_{photo}$ and $\mid$$\Delta \mu_{dot}$$\mid $ also start to saturate at high power.  We found a very robust nonlinear dependency between $\Delta \mu$, $I_{photo}$ and the incident power for all $V_{g}$. These tendencies are not compatible with a $\Delta \mu_{dot}$ induced by the intraband absorption of the THz photons in the graphene leads since at such low incident THz electric field and low temperature \cite{Hafez2018}, $\Delta \mu_{dot}$ is expected to evolve linearly with $P_{0}$ (see Supporting Information). On the contrary, both the nonlinearity and saturation features are consistent with interfacial photogating effect. Indeed, as trapped photocarrier lifetime in silicon depends on the incident optical power \cite{Vinh2013}, the photoexcited carrier density  in the steady-state regime contributing to the photogating effect is expected to evolve nonlinearly with $P_{0}$. The observed saturation can be attributed to a decrease of the height of the potential well at the SiO$_{2}$/Si interface as the THz incident power is increased. This feature has been observed in several previous works in graphene-based photodetectors dominated by the photogating effect \cite{Luo2018} and attributed to the trapped charges at the interface that induce an opposite build-in electric field. Thus, these power-resolved measurements provide an additional evidence that interfacial photogating is the main physical mechanism involved in the THz photoresponse of the GQD-based device.

In conclusion, we have investigated a hBN-encapsulated GQD in Coulomb blockade regime under THz illumination at low temperature. The hBN encapsulation enables to fabricate low roughness GQD with excited states at energy $\sim$ 1.7 meV (i.e. 0.4 THz) directly observable by transport experiments. We demonstrate a large photocurrent, in the nanoampere range, under incoherent THz illumination originating from a renormalization of the chemical potential of the GQD with respect to those of the leads. We show that this effect evolves nonlinearly with the incident power and attribute it to interfacial photogating effect. Besides, we evidence the absence of thermal effects in the Coulomb blockade regime, paving the way for probing photon assisted tunneling transport by exciting the GQD device with coherent THz photons \cite{Kawano2008} and THz light-matter coupling by the insertion of the GQD in a THz resonator. Furthermore, owing to their great flexibility in electronic states engineering, these large GQDs are very promising for the developments of THz emitters and detectors.


\begin{acknowledgement}
The authors thank Bernard Pla\c cais for valuable discussions. 
\textbf{Funding}: This project has received funding from the European Research Council (ERC) under the European Union’s Horizon 2020 research and innovation program (grant agreement No. 820133).
K.W. and T.T. acknowledge support from the Elemental Strategy Initiative conducted by the MEXT, Japan ,Grant Number JPMXP0112101001,  JSPS KAKENHI Grant Numbers JP20H00354 and the CREST(JPMJCR15F3). 
\textbf{Author contributions}: E.R. fabricated the sample and acquired and interpreted the experimental data. S.B, F.V. set up the cryogenic experiment. M.D. and T.K. set up the transport experiment aquisition. S.M. K.W. and T.T. participated in sample fabrication.  
J.M., T.K., R.F interpreted the experimental data. S.M. contributed to THz power and absorption estimation, heat diffusion calculations and antenna FEM simulation. The manuscript was written and the data were interpreted by E.R., T.K., R.F., S.B., J.M., S.D., J.T.  and T.K. provided insights into manuscript writing. All work was coordinated and overseen by J.M. All authors contributed to the discussion and to the final preparation of the manuscript.
\textbf{Competing interests}: The authors declare that they have no competing interest.
\textbf{Data and materials availability}: All data needed to evaluate the conclusions in the paper are present in the paper and/or the Supporting Information. Additional data related to this paper may be requested from the authors.

\end{acknowledgement}


\section{Supporting Information}
The supporting information is available free of charge at 
https://pubs.acs.org/doi/10.1021/acs.nanolett.0c01800

\bibliography{Riccardi_bibl}

\end{document}